\documentclass{iau}
\usepackage{graphicx}

\title[Solar-like oscillations in subgiant and red-giant stars] 
{Solar-like oscillations in subgiant and red-giant stars: mixed modes}

\author[S. Hekker \& A. Mazumdar]   
{S. Hekker$^{1,2}$ \and A. Mazumdar$^3$}

\affiliation{$^1$ Max Planck Institute for Solar System Research, Katlenburg-Lindau, Germany\\ email: {\tt Hekker@mps.mpg.de} \\[\affilskip]
$^2$ Astronomical Institute ``Anton Pannekoek'', University of Amsterdam, Amsterdam, the Netherlands\\[\affilskip]
$^3$ Homi Bhabha Centre for Science Education, TIFR, Mumbai, India}

\pubyear{2014}
\volume{301}  
\pagerange{1--8}
\jname{Precision Asteroseismology}
\editors{J.A. Guzik, W.J. Chaplin, G. Handler \& A. Pigulski, eds.}
\begin{document}

\maketitle

\begin{abstract}
Thanks to significant improvements in high-resolution spectrographs and the launch of dedicated space missions \textit{MOST}, \textit{CoRoT} and \textit{Kepler}, the number of subgiants and red-giant stars with detected oscillations has increased significantly over the last decade. The amount of detail that can now be resolved in the oscillation patterns does allow for in-depth investigations of the internal structures of these stars. One phenomenon that plays an important role in such studies are mixed modes. These are modes that carry information of the inner radiative region as well as from the convective outer part of the star allowing to probe different depths of the stars.

Here, we describe mixed modes and highlight some recent results obtained using mixed modes observed in subgiants and red-giant stars. 
\end{abstract}

\firstsection 
\section{Introduction}

Solar-like oscillations are oscillations stochastically excited in the
outer convective layers of low-mass stars on the main-sequence (such
as the Sun), subgiants and red-giant stars (e.g. \cite[Goldreich \&
  Keeley 1977]{goldreich1977}, \cite[Goldreich \& Kumar
  1988]{goldreich1988}). Effectively, some of the convective energy is
converted into energy of global oscillations. The main characteristics
of these oscillations are that a) essentially all modes are excited
albeit with different amplitudes; b) the stochastic driving and
damping results in a finite mode lifetime.

\begin{figure}
\centering
\includegraphics[width=\linewidth]{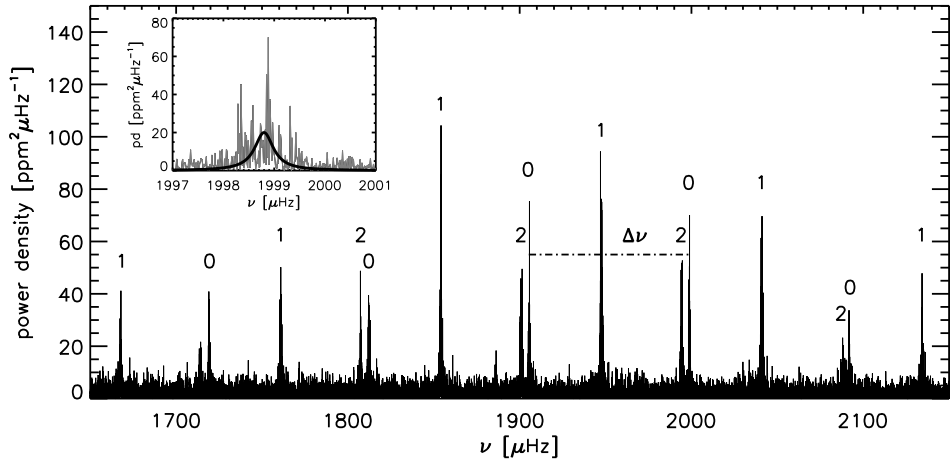}
\caption{Fourier power density spectrum of a main-sequence star (KIC 3656479, \cite[Hekker 2013]{hekker2013}). The angular degree ($\ell$) of the modes are indicated. The horizontal dashed-dotted line indicates $\Delta\nu$ (see Eq.~\ref{tassoul}). The inset shows a zoom of the radial mode at $\sim$1999 $\mu$Hz with a Lorentzian profile overplotted.}
\label{ps}
\end{figure}

In Fourier space the solar-like oscillation characteristics yield a
regular pattern of oscillation frequencies with a roughly Gaussian
envelope, where each frequency peak has a width inversely proportional
to the mode lifetime. Hence, resolved individual
frequency peaks can be fitted using a Lorentzian profile to determine
height, width and frequency of the oscillation mode (see inset in
Fig.~\ref{ps}). The regular pattern of the frequencies in the Fourier
spectrum can be described asymptotically (\cite[Tassoul
  1980]{tassoul1980}, \cite[Gough 1986]{gough1986}):

\begin{equation}
\nu_{n,\ell} \simeq \Delta\nu\left(n+\frac{\ell}{2}+\epsilon\right)-\delta\nu_{n,\ell},
\label{tassoul}
\end{equation}
where $\nu$ is frequency and $\epsilon$ is a phase term. $\Delta\nu$ is the regular spacing in frequency of oscillations with the same angular degree $\ell$ and consecutive orders $n$ (large frequency separation, horizontal dashed-dotted line in Fig.~\ref{ps}). $\delta\nu_{n,\ell}$ is the so-called small frequency separation between pairs of odd or even modes.
The large frequency separation, $\Delta\nu$, is proportional to the square root of the mean density of the star. Another characteristic of the Fourier spectrum is the frequency of maximum oscillation power, $\nu_{\rm max}$, which depends on the surface gravity and effective temperature of the star. 

For a full overview of solar-like oscillations we refer to \cite[Aerts et al.~(2010)]{aerts2010}, \cite[Chaplin \& Miglio (2013)]{chaplin2013} and \cite[Hekker (2013)]{hekker2013}.

\section{Mixed modes}
\subsection{Cavities}

\begin{figure}
\centering
\includegraphics[width=0.7\linewidth]{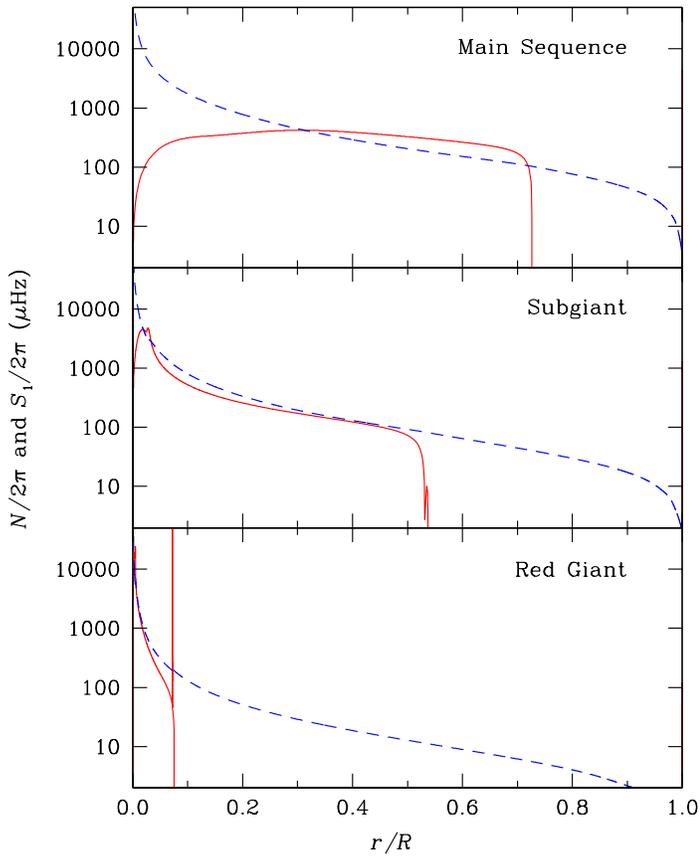}
\caption{The Brunt-V\"ais\"al\"a ($N$, red solid line) and the Lamb frequency for $\ell=1$ modes ($S_1$, blue dashed line) are shown as functions of fractional radius ($r/R$) for models of a $1\,M_{\odot}$ star in the main sequence (top panel), subgiant (middle panel) and red-giant (bottom panel) phases.}
\label{cavs}
\end{figure}

The oscillations described above are pure acoustic pressure (p)
modes. Non-radial p-modes reside in a cavity in the outer parts of the
star bound at the bottom by the Lamb frequency ($S_{\ell}$), where the
horizontal phase speed of the wave equals the local sound speed. At
the top the cavity is limited by the cut-off frequency $\nu_{\rm ac} \propto
g / \sqrt{T_{\rm eff}}$ (\cite[Brown et al.~1991]{brown1991}) above
which the atmosphere is not able to trap the modes and the
oscillations become traveling waves, so called high-frequency or
pseudo-modes (e.g., \cite[Karoff 2007]{karoff2007}). When stars evolve
the p-mode frequencies decrease, mostly due to the decrease in surface
gravity (increase in radius) and hence decrease in cut-off
frequency. At the same time oscillations that reside in the inner
radiative region of the star have increasing frequencies with
evolution. These so-called gravity (g) modes have buoyancy as their
restoring force. These modes reside in a cavity defined by
finite values of the Brunt-V\"ais\"al\"a or buoyancy frequency
$N$. This is the frequency at which a vertically displaced parcel will
oscillate within a statically stable environment. The peak of the
Brunt-V\"ais\"al\"a frequency increases with evolution due to the
increase in the core gravity. Oscillations can only be sustained when
$\nu < S_{\ell},N$ (g mode) or $\nu > S_{\ell},N$ (p mode). A region
where either of these conditions is not satisfied is an evanescent
zone for the respective mode at a particular frequency. See
Fig.~\ref{cavs} for the dipole Lamb frequency and Brunt-V\"ais\"al\"a
frequency as a function of stellar radius defining the respective p-
and g-mode cavities of a main-sequence star (top panel), subgiant
(middle panel) and a red giant (bottom panel).

\subsection{Avoided crossings}
In subgiants and red giants the frequencies in both p- and g-mode cavities have similar values and a coupling between these frequencies can persist if the evanescent zone is narrow, i.e., the oscillation is not damped out. In that case a p and g mode with similar frequencies and same spherical degree undergo a so-called avoided crossing. The interactions between the modes will affect (or bump) the frequencies. This bumping can be described as a resonance interaction of two oscillators (e.g. \cite[Aizenman et al.~1977]{aizenman1977}, \cite[Deheuvels \& Michel 2010]{deheuvels2010}, \cite[Benomar et al.~2013]{benomar2013}).

In short the avoided crossings can be viewed  using a system of two coupled oscillators $y_1(t)$ and $y_2(t)$ with a time dependence:
\begin{equation}
\frac{d^2y_1(t)}{dt^2} = -\omega_1^2y_1 + \alpha_{1,2}y_2
\end{equation}
\begin{equation}
\frac{d^2y_2(t)}{dt^2} = -\omega_2^2y_1 + \alpha_{1,2}y_1
\end{equation}
where $\alpha_{1,2}$ is the coupling term between the two oscillators, and  $\omega_1$, $\omega_2$ are the eigenfrequencies of the uncoupled oscillators ($\alpha_{1,2} = 0$). In the case of uncoupled oscillators the eigenfrequencies can cross at $\omega_0$ where $\omega_1=\omega_2$.

\begin{figure}
\centering
\includegraphics[width=\linewidth]{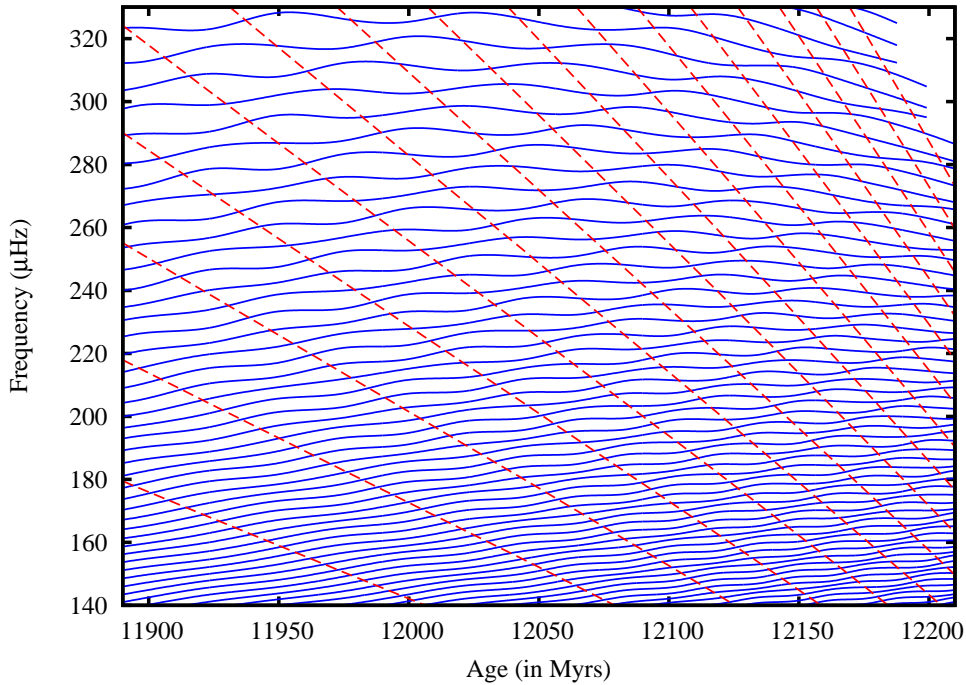}
\caption{The evolution of frequencies of a $1M_{\odot}$ star with age in the red-giant phase. The blue continuous lines depict the $\ell=1$ modes while the red dashed lines represent the $\ell=0$ modes of different radial orders.}
\label{bump}
\end{figure}

\begin{figure}
\centering
\includegraphics[width=\linewidth]{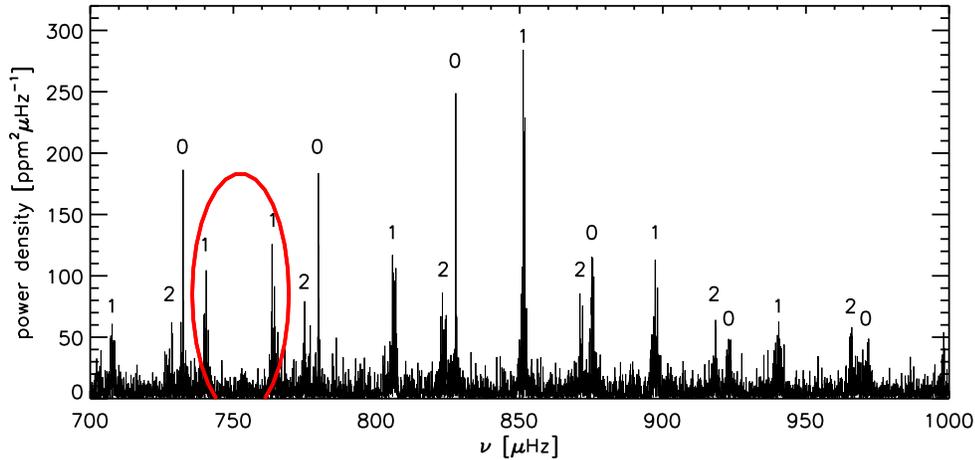}
\caption{Fourier power density spectrum of a subgiant (KIC 11395018, \cite[Hekker 2013]{hekker2013}). The degree ($\ell$) of the modes is indicated. A mixed mode pair is present at 740.3 and 764.0~$\mu$Hz highlighted by the (red) oval.}
\label{pssubgiant}
\end{figure}

If the coupling term $\alpha_{1,2}$ is very small compared to the difference between the eigenfrequencies ($\alpha_{1,2} \ll |\omega_1^2 - \omega_2^2|$), then the eigenfrequencies of the system are hardly perturbed and have values close to $\omega_1$ and $\omega_2$. However, if the difference between the eigenfunctions is small compared to the coupling term ($|\omega_1^2 - \omega_2^2| \ll \alpha_{1,2}$), then the eigenfrequencies can be approximated by $\omega^2 = \omega_0^2 \pm \alpha_{1,2}$. These two eigenmodes behave as mixed modes, one with dominant features from $\omega_1$ and the other one with dominant features from $\omega_2$. 

For a $1M_{\odot}$ star, Fig.~\ref{bump} shows the variation of frequencies with evolution. The general trend of the p-mode frequencies is to decrease with age, while g-mode frequencies increase with age (see section 2.1). Therefore at a particular age, a g mode and a p mode of the same angular degree and similar frequencies can exist, which would --- instead of crossing each other to continue these opposite trends --- interact  to produce a pair of mixed modes with close frequencies, as explained above. This is visible in Fig.~\ref{bump} as a series of bumps in the $\ell =1$ modes, where each bump is located at an avoided crossing. One of these mixed modes will have dominant features from the underlying p mode, while the other mixed mode will have dominant features from the underlying g mode. Fig.~\ref{pssubgiant} shows a Fourier power density spectrum of a subgiant with a mixed-mode pair.

As evident from Fig.~\ref{bump} an avoided crossing can occur at a specific frequency only at a specific age. Thus observations of mixed modes allows for a precise estimate of the age. However, the age determination is model dependent and the actual value of the stellar age might differ as a function of physical processes included in the model.

\begin{figure}
\centering
\includegraphics[width=\linewidth]{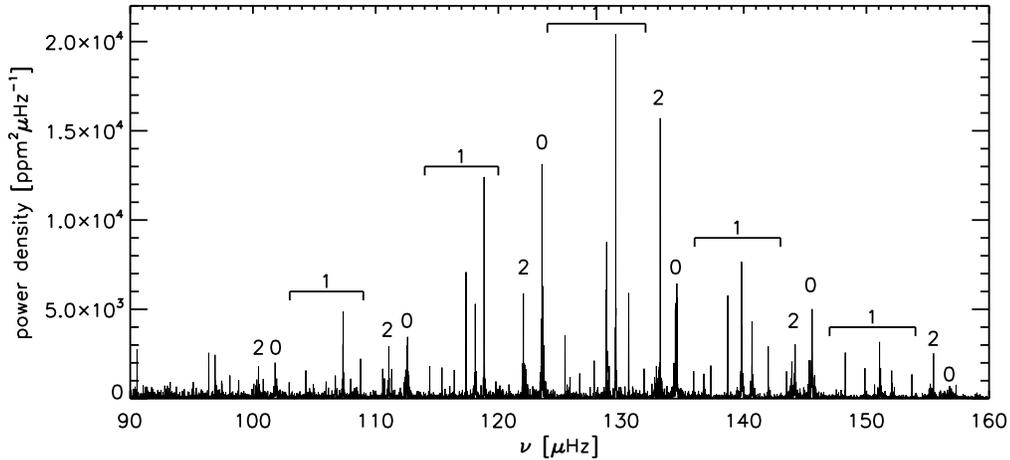}
\caption{Fourier power density spectrum of a red-giant star (KIC 9145955, \cite[Hekker 2013]{hekker2013}). The degree ($\ell$) of the modes is indicated. For the dipole modes the approximate range of the observed mixed modes is indicated.}
\label{psgiant}
\end{figure}

In more evolved stars the density of g modes around a given frequency can be high and multiple g modes can interact with different coupling terms with a single p mode to produce multiple mixed modes. Fig.~\ref{psgiant} shows a Fourier power density spectrum of a red-giant star with multiple mixed dipole modes. 

\subsection{Period spacing}
The high-order g modes in the inner radiative region have (in an asymptotic approximation) a typical spacing in period ($\Delta\Pi$, e.g. \cite[Tassoul 1980]{tassoul1980}).  Since the regular pattern of the frequencies is broken by the avoided crossings, the period spacing will also be affected. For red giants with multiple mixed modes per p-mode order, this results in smaller observed $\Delta\Pi$  for pressure-dominated mixed modes (lying close to the underlying p mode) and increasing values of $\Delta\Pi$ for mixed modes with more g-dominated character, i.e., with frequencies further away from the underlying p mode. The least coupled, i.e. most g-dominated, modes have a $\Delta\Pi$ close to the asymptotic value. However these modes are the hardest to detect due to their low heights in the Fourier power spectrum (\cite[Dupret et al.~2009]{dupret 2009}).

\section{Recent results}

\subsection{Subgiants}
For subgiants a recent highlight is the detection of radial differential rotation \cite[(Deheuvels et al. 2012)]{deheuvels2012}.  Due to the different sensitivities of mixed modes to different parts of the star (either the internal or the outer region of the star depending on their predominant g- or p-mode nature) it is possible to study stellar properties, such as rotation, as a function of radius.  \cite[Deheuvels et al.~(2012)]{deheuvels2012} were able to detect rotational splittings in 17 $\ell =1$ mixed modes in a subgiant, which have different pressure-gravity mode sensitivity. They concluded that  the core rotates approximately five times faster than the surface.

\subsection{Red giants}
Gravity-dominated mixed modes in red-giant stars have recently been discovered observationally for $\ell = 1$ modes by \cite[Beck et al.~(2011)]{beck2011}. From these mixed modes the period spacings can be derived. \cite[Bedding et al.~(2011)]{bedding2011} and \cite[Mosser et al.~(2011)]{mosser2011mm} showed that the observed period spacings of red giants in the hydrogen-shell-burning phase ascending the red-giant branch are significantly different from the period spacing for red giants also burning helium in the core. This provides a clear separation into these two groups of stars that are superficially very similar. This effect was explained by \cite[Christensen-Dalsgaard (2011)]{jcd2011} as being due to convection in the central regions of the core in the helium-burning stars. The buoyancy frequency is nearly zero in the convective region. As the period spacing is inversely proportional to the integral over $N$, the period spacing of stars with convection in the core is higher compared to the period spacing of stars without a convective region in the core.

Similar to the subgiant mentioned in the previous subsection, the rotational splittings of dipole mixed modes with different p/g nature have been measured for a red-giant star (\cite[Beck et al.~2012]{beck2012}), which revealed that the core rotates approximately ten times faster than the surface. 
Subsequently, \cite[Mosser et al.~(2012)]{mosser2012rot} showed that stars ascending the red-giant branch experience a small increase of the core rotation followed by a significant slow-down in the later stages of the red-giant branch resulting in slower rotating cores in red-clump (or horizontal-branch) stars compared to faster rotating cores in stars on the red-giant branch. 

\section{Future}
With the long term datasets currently available from \textit{Kepler}, mixed modes can be detected in many stars. This allows for further studies of the radial differential rotation (see Section 3 and Di Mauro et al., these proceedings). Furthermore, mixed modes will also allow for further studies of the internal structure of subgiants and red-giant stars, possibly including core overshoot, and the presence of secondary Helium flashes. The latter are present in stellar evolution models of low-mass stars which ignite Helium in a degenerate core, but it is still unclear whether this is a realistic representation. Therefore, it is evident that  mixed modes have great diagnostic potential especially for subgiants and red-giant stars.

\begin{acknowledgements}
SH acknowledges financial support from the Netherlands Organisation for Scientific Research (NWO). AM acknowledges support from the NIUS programme of HBCSE.
\end{acknowledgements}


\begin{thebibliography}{}

\bibitem[{{Aerts} et~al.(2010){Aerts}, {Christensen-Dalsgaard},
  and {Kurtz}}]{aerts2010}
{Aerts}, C., {Christensen-Dalsgaard}, J., {Kurtz}, D.W., 2010, \textit{Asteroseismology, Astronomy and Astrophysics Library.~ISBN 978-1-4020-5178-4.~Springer Science}

\bibitem[{{Aizenman} et~al.(1977){Aizenman}, {Smeyers}, and
  {Weigert}}]{aizenman1977}
{Aizenman}, M., {Smeyers}, P., \& {Weigert}, A. 1977, \textit{A\&A}, 58, 41
  
 \bibitem[{{Beck} et~al.(2011){Beck}, {Bedding}, {Mosser}, and {et
  al.}}]{beck2011}
{Beck}, P.G., {Bedding}, T.R., {Mosser}, B., {et al.} 2011, \textit{Science}, 332, 205

\bibitem[{{Beck} et~al.(2012){Beck}, {Montalban}, {Kallinger}, and {et
  al.}}]{beck2012}
{Beck}, P.G., {Montalb\'an}, J., {Kallinger}, T., {et al.} 2012, \textit{Nature}, 481, 55

\bibitem[{{Bedding} et~al.(2011){Bedding}, {Mosser}, {Huber}, and {et
  al.}}]{bedding2011}
{Bedding}, T.R., {Mosser}, B., {Huber}, D., {et al.} 2011, \textit{Nature}, 471, 608
  
\bibitem[{{Benomar} et~al.(2013){Benomar}, {Bedding}, {Mosser}, and {et
  al.}}]{benomar2013}
{Benomar}, O., {Bedding}, T.R., {Mosser}, B., {et al.} 2013, \textit{ApJ}, 767, 158

\bibitem[{{Chaplin} and {Miglio}(2013)}]{chaplin2013}
{Chaplin}, W.J., \& {Miglio}, A. 2013, \textit{ARA\&A}, 51, 353

\bibitem[{{Christensen-Dalsgaard}(2011)}]{jcd2011}
{Christensen-Dalsgaard}, J. 2011, \textit{arXiv}: 1106.5946

\bibitem[{{Deheuvels} et~al.(2012){Deheuvels}, {Garc{\'{\i}}a}, {Chaplin}, and
  {et al.}}]{deheuvels2012}
{Deheuvels}, S., {Garc{\'{\i}}a}, R.A., {Chaplin}, W.J., {et al.} 2012,
\textit{ApJ}, 756, 19

\bibitem[{{Deheuvels} and {Michel}(2010)}]{deheuvels2010}
{Deheuvels}, S., \& {Michel}, E. 2010, \textit{Ap\&SS}, 328,
  259
  
 \bibitem[{{Dupret} et~al.(2009){Dupret}, {Belkacem}, {Samadi}, and {et
  al.}}]{dupret2009}
{Dupret}, M.-A., {Belkacem}, K., {Samadi}, R., {et al.} 2009,
\textit{A\&A}, 506, 57

\bibitem[{{Goldreich} and {Keeley}(1977)}]{goldreich1977}
{Goldreich}, P., \& {Keeley}, D.A. 1977, \textit{ApJ}, 212,
  243

\bibitem[{{Goldreich} and {Kumar}(1988)}]{goldreich1988}
{Goldreich}, P., \& {Kumar}, P. 1988, \textit{ApJ}, 326, 462

\bibitem[{{Gough}(1986)}]{gough1986}
{Gough}, D.O. 1986, \textit{Highlights
  of Astronomy}, 7, 283

 \bibitem[{{Hekker}(2013)}]{hekker2013}
{Hekker}, S. 2013, \textit{Advances in Space Research}, 52, 1581

\bibitem[{{Karoff}(2007)}]{karoff2007}
{Karoff}, C. 2007, \textit{MNRAS}, 381, 1001

\bibitem[{{Mosser} et~al.(2011){Mosser}, {Barban},
  {Montalb{\'a}n}, and {et al.}}]{mosser2011mm}
{Mosser}, B., {Barban}, C., {Montalb{\'a}n}, J., {et al.}
  2011, \textit{A\&A}, 532, A86
  
 \bibitem[{{Mosser} et~al.(2012){Mosser}, {Goupil}, {Belkacem}, and {et
  al.}}]{mosser2012rot}
{Mosser}, B., {Goupil}, M.J., {Belkacem}, K., {et al.} 2012, \textit{A\&A}, 548, A10

 \bibitem[{{Tassoul}(1980)}]{tassoul1980}
{Tassoul}, M. 1980, \textit{ApJS}, 43, 469
\end{thebibliography}
\end{document}